# Dramatic Mobility Enhancements in Doped SrTiO$_3$ Thin Films by Defect Management

Y. Kozuka,[1,a)] Y. Hikita,[1] C. Bell,[1,2] and H. Y. Hwang [1,2]

[1]*Department of Advanced Materials Science, University of Tokyo, Chiba 277-8561, Japan*

[2]*Japan Science and Technology Agency, Kawaguchi, Saitama 332-0012, Japan*

We report bulk-quality *n*-type SrTiO$_3$ (*n*-SrTiO$_3$) thin films fabricated by pulsed laser deposition, with electron mobility as high as 6600 cm$^2$ V$^{-1}$ s$^{-1}$ at 2 K and carrier density as low as $2.0 \times 10^{18}$ cm$^{-3}$ (~ 0.02 at. %), far exceeding previous pulsed laser deposition films. This result stems from precise strontium and oxygen vacancy defect chemistry management, providing a general approach for defect control in complex oxide heteroepitaxy.

[a)]Present address: Institute for Materials Research, Tohoku University, Sendai 980-8577, Japan; electronic mail: kozuka@imr.tohoku.ac.jp

SrTiO$_3$ has long been used as a substrate for growing high-quality perovskite-related oxide films. It is also a fascinating material in its own right, exhibiting quantum paraelectricity,[1] metallic behavior,[2,3] and superconductivity[4] at semiconductor densities far below that of other 3*d* transition-metal perovskites. Increasingly, these features are being incorporated at heterointerfaces[5-10] and in devices.[11-15] Successful efforts thus far have generally been restricted to interfaces with bulk substrates. This limitation arises from the common degradation of thin film SrTiO$_3$ physical properties. For example, the low-temperature dielectric constant of typical SrTiO$_3$ films is ~ 1000:[16] far below the bulk value of ~ 20000.[1] The low-temperature electron mobility can exceed 10000 cm$^2$ V$^{-1}$ s$^{-1}$ for bulk doped SrTiO$_3$,[2,3] while, for typical substitutionally-doped films grown by pulsed laser deposition, values are a few hundred cm$^2$ V$^{-1}$ s$^{-1}$ at maximum,[17-19] with problematic carrier deactivation in many cases.[20] Improving these limitations would greatly enhance the use of thin film SrTiO$_3$ as a doped semiconductor or superconductor building block in complex heterostructures.

The central issue for growing semiconductor-device-quality SrTiO$_3$ thin films is strict defect management,[21] due to the low dopant density (typically < 0.1 at. %), in contrast to other functional transition-metal oxides, which are heavily doped (typically > 10 at. %). To achieve this, we note that in conventional semiconductor films, quality is significantly improved by using a growth temperature above half the bulk melting point.[22] Above this point, enhanced surface adatom migration improves the crystallinity, as indicated by significant dielectric enhancements in the case of SrTiO$_3$.[23] Furthermore, in these conditions the defect densities tend towards equilibrium values, which are calculable.[21] Here we apply these concepts to SrTiO$_3$, resulting in a dramatic improvement in low-temperature electron mobility.

A first-principles calculation showed that the Sr partial Schottky defect is energetically favored over the Ti partial Schottky defect.[24] The formation equilibrium of the former is expressed as $Sr_{Sr}^{\times} + O_O^{\times} \leftrightarrow V_{Sr}'' + V_O^{\bullet\bullet} + SrO_{R/P}$, in the notation of Kröger and Vink,[25] where $SrO_{R/P}$ denotes the Ruddlesden-Popper phase. The corresponding mass-action law (low oxygen partial pressure limit) is reduced to

$$[V_{Sr}''][V_O^{\bullet\bullet}] = K_S^{\circ} \exp\left(-\frac{E_S}{kT}\right), \qquad (1)$$

where $k$ is Boltzmann's constant, and $T$ temperature. $K_S^{\circ}$ and $E_S$ are constants, experimentally determined in Ref. 21. Equation (1) indicates for a low density of oxygen vacancies, a high-density of strontium vacancies is formed, as shown in the defect density calculation (Fig. 1). In particular, at high oxygen partial pressure, electrons supplied by extrinsic Nb (or La) donors are completely compensated by doubly-charged Sr vacancies (Fig. 1(b), corresponding to $[V_{Sr}''] = [Nb]/2$). This is the likely origin of carrier deactivation in high pressure grown films. Therefore, enhancing the formation of $V_O^{\bullet\bullet}$ using low oxygen pressure growth, $V_{Sr}''$ can be greatly suppressed.

Subsequent oxygen annealing easily eliminates oxygen vacancies. However, care is needed to avoid regenerating Sr vacancies during this post-annealing.[26] The several orders of magnitude larger diffusion constant of oxygen vacancies compared to cation vacancies in SrTiO$_3$ can be utilized.[27-29] Indeed, it was reported that strontium vacancies are not mobile in a reasonable time scale below 1000 ºC,[21] and oxygen vacancies can be mobile at 400 ºC. Thus high temperature growth in reducing conditions, followed by a lower temperature oxidizing anneal, utilizes the defect thermodynamic equilibrium to greatly improve film quality.

We deposited $n$-SrTiO$_3$ by pulsed laser deposition (PLD), using an ultrahigh vacuum chamber with an infrared laser heating system,[30] enabling sample heating to 1300 ºC, as measured by an optical pyrometer. Ablation targets were commercial $n$-SrTiO$_3$ single crystals with at. % dopant concentrations of 5, 1, 0.2, 0.1, and 0.02. La is the dopant for 5 at. %, while Nb was used elsewhere. Substrates were undoped SrTiO$_3$ (100) single crystals (Shinkosha Co.), treated by buffered HF to remove the mechanical polishing surface damage.[31] During growth, the oxygen partial pressure was less than $10^{-7}$ Torr. For ablation, a 248 nm KrF excimer laser was used at a fixed frequency of 5 Hz. The laser fluence on the target surface was 0.5 J/cm$^2$ (spot size = 0.04 cm$^2$). The substrate temperature was changed at a constant rate of 30 ºC/minute. The samples were postannealed either *in situ* at 900 ºC (one hour, fixed oxygen pressure = $10^{-2}$ Torr), or *ex situ* at 600 ºC (24 hours, flowing oxygen) to fill oxygen vacancies. Both postannealing conditions eliminated all oxygen vacancies, giving no contribution to electrical conduction. The thin films were characterized by x-ray diffraction to extract the out-of-plane lattice constant. Resistivity and Hall measurements were made using a standard four-point method.

First, the lattice constant vs growth temperature relationship was investigated using 100 nm-thick 0.1 at. % Nb-doped SrTiO$_3$ thin films deposited on SrTiO$_3$ substrates. Lower growth temperatures (< 1050 ºC) gave lattice elongation from the bulk value (0.3905 nm), as shown in Fig. 2. In addition, the sheet resistance was > 100 MΩ/sq. at 295 K. These features can be attributed to cation off-stoichiometry, as found previously.[32] However, above 1050 ºC the lattice constant closely matched the bulk value. Metallic conduction was observed, as indicated by the 2 K electron mobility (Fig. 2). This abrupt transition at 1050 ºC is close to half of the melting point of SrTiO$_3$ (~ 2000 ºC),[33] corresponding well with the threshold found for improved semiconductor growth,[22] which was suggested to be the surface quasi-melting point. The observed transition in the lattice constant supports this notion, since in the SrTiO$_3$ phase diagram the congruent point matches the stoichiometric point.[33] This assures macroscopically homogeneous film stoichiometry in the equilibrium state: indeed high growth temperature films were rather insensitive to plume stoichiometry, unlike lower temperature growth.[32]

Given the high-temperature deposition, the extent of Nb diffusion was measured by secondary ion mass spectroscopy. For this purpose, the structure SrTiO$_3$/0.1 at. % Nb-doped SrTiO$_3$/SrTiO$_3$ was fabricated at 1200 ºC on a 0.1 at. % Nb-doped SrTiO$_3$ substrate. The substrate was used as the Nb density standard (nominal value $=1.68\times10^{19}$ cm$^{-3}$). The data (Fig. 3) indicate that the scale of Nb diffusion is less than the measurement's spatial resolution (~ 5 nm). The asymmetry between the top and bottom interfaces of the film is caused by the knock-on effect: the sputtering Cs ions push the Nb ions into the surface. We conclude a negligible contribution of Nb diffusion to the conduction despite the high growth temperature.

We next deposited 100 nm-thick *n*-SrTiO$_3$ thin films using the five targets with at. % dopant concentrations of 5, 1, 0.2, 0.1, and 0.02. The growth temperature was 1200 ºC which was optimal for mobility (Fig. 2). Atomic force microscopy of films grown at 1300 ºC, showed significant surface roughening, while a clear atomic step-and-terrace surface was found for films grown up to 1200 ºC (not shown). Metallic conductivity was found for all films except for 0.02 at. % Nb-doped SrTiO$_3$ thin films, which was insulating due to surface depletion.[34] The estimated surface depletion width is 280 nm, (assuming a 0.7 V surface pinning potential, 20000 dielectric constant, and $2.0 \times 10^{18}$ cm$^{-3}$ carrier density): exceeding the film thickness. By increasing the thickness to 500 nm, metallic 0.02 at. % Nb-doped SrTiO$_3$ could also be formed.

Hall measurements indicated an almost constant carrier density, while the electron mobility monotonically increased, from 300 K to 2 K, [Figs. 4(a) and 4(b)]. At 2 K, we obtained a maximum electron mobility of 6600 cm$^2$ V$^{-1}$ s$^{-1}$ (carrier density $2.0 \times 10^{18}$ cm$^{-3}$, 0.02 at. %). The excellent agreement between the nominal carrier density and the measured density over many orders of magnitude, [Fig. 4(c)] indicates negligible oxygen vacancy electrical conductivity. The electron mobility of the films at 2 K is comparable to the previous best reports of bulk single crystals values at all densities, [Fig. 4(d)].

Finally, we note that our growth regime mimics that used for bulk *n*-SrTiO$_3$ single crystal synthesis. After the highly non-equilibrium conditions of Verneuil growth, crystals are usually annealed in moderately reducing conditions, to suppress compensation of donors by cation vacancies, and reduce strain.[35] Although nanoscale spatial modulation of oxygen vacancies has been demonstrated in PLD SrTiO$_3$ films,[36] the very narrow growth window for oxygen vacancy doping[37] precludes the reproducible growth at the densities demonstrated here. With this robust method of controlling substitutional doping, we can now access a far wider phase space, opening up the possibility of combining high electron mobility, low carrier density and superconductivity in SrTiO$_3$ heterostructures.[38]

We acknowledge MST Co. (Japan) for the secondary ion mass spectroscopy measurement, and T. Ohnishi and M. Lippmaa for helpful discussions.

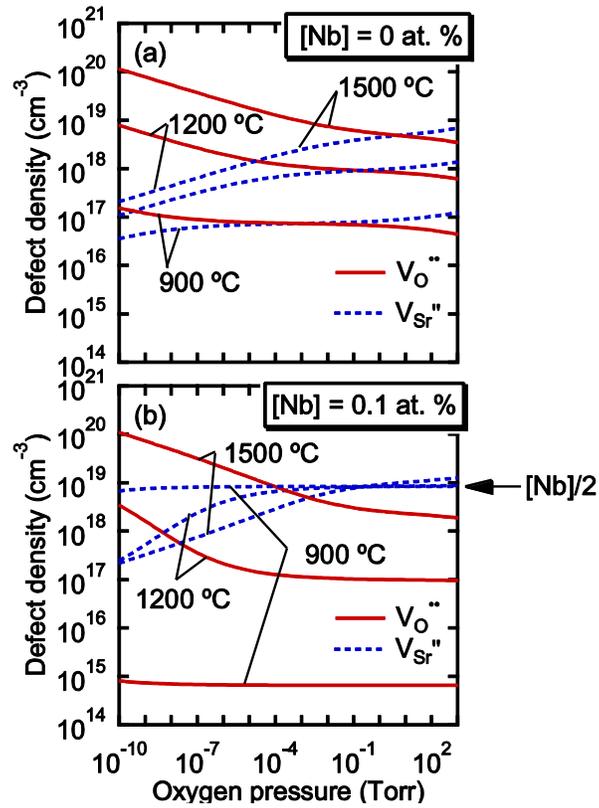

FIG. 1. (Color online) Calculated Sr and oxygen vacancy concentrations vs oxygen partial pressure at 900 ºC, 1200 ºC, and 1500 ºC following Ref. 21. Doping concentrations are (a) Nb = 0 at. % and (b) Nb = 0.1 at. %. [Nb]/2 is also shown in (b).

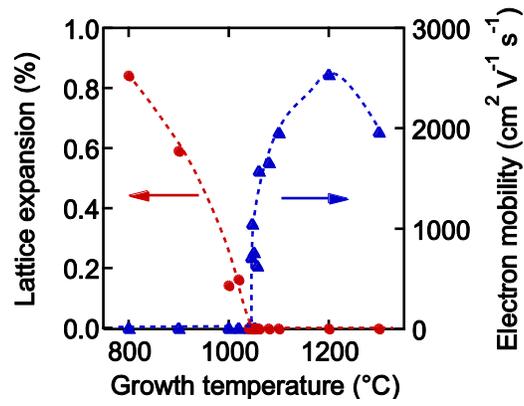

FIG. 2. (Color online) Expansion of the out-of-plane lattice constant from the bulk value (0.3905 nm) and electron mobility vs growth temperature for Nb = 0.1 at. % thin films. The lattice constant was obtained from room temperature x-ray diffraction. The electron mobility was estimated at 2 K. Dashed curves are guides to the eye.

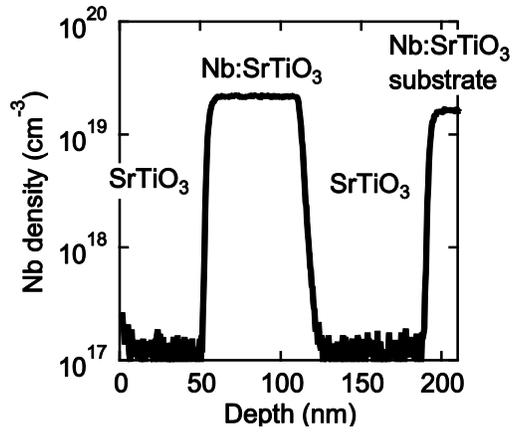

FIG. 3. Nb density profile measured by secondary ion mass spectroscopy, normalized by the substrate nominal value, for the structure: $SrTiO_3$(50 nm)/0.1 at. % Nb-doped $SrTiO_3$(50 nm)/$SrTiO_3$(70 nm)/0.1 at. % Nb-doped $SrTiO_3$ substrate.

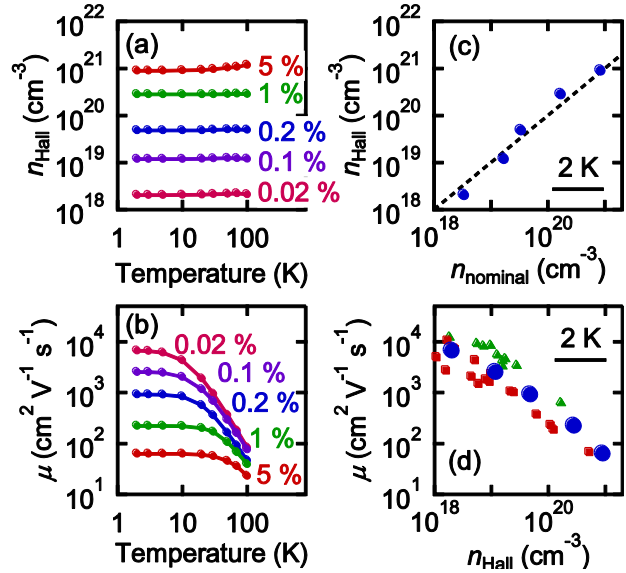

FIG. 4. (Color online) (a) Carrier density and (b) electron mobility of $n$-SrTiO$_3$ thin films vs temperature for five dopant concentrations, estimated from the Hall effect. (c) Nominal dopant density vs measured values at 2 K. (d) Electron mobility of the films vs carrier density (circles), compared to oxygen deficient (squares) and Nb-doped (triangles) bulk single crystals (Refs. 2,3). The thickness is 500 nm for the 0.02 at. % film, 100 nm otherwise.

13